# The Double Cluster G185 in M31

Stephen Holland, Gregory G. Fahlman[1], & Harvey B. Richer
Department of Geophysics & Astronomy,
University of British Columbia,
Vancouver, B.C.,
V6T 1Z4

## ABSTRACT

We have identified a small globular cluster in M31 located approximately $4''$ northwest of the M31 globular cluster G185. While several multiple globular clusters have been observed in the Magellanic Clouds none have been found in the Galaxy or in M31. We estimate the probability of such a chance line-of-sight alignment occuring near the nucleus of M31 to be $0.09 \pm 0.03$ and find no obvious indication of any tidal deformation in either cluster, as would be expected if the clusters were interacting.

Two-dimensional modelling suggests G185 has a King (1966) [AJ, 71, 64] concentration of $c = 1.11 \pm 0.08$ while the companion has $c = 0.67 \pm 0.17$ and is physically smaller than G185. Both objects have integrated dereddened colors similar to those of Galactic globular clusters.

## 1. Introduction

Binary globular star clusters are dynamically very fragile objects so it is not surprising that none have been observed in the Galaxy. Innanen et al. (1972) have used numerical simulations to show that a binary globular cluster with a separation of less than ∼50 pc and a periGalacticon of less than ∼1 kpc would not survive a single Galactic orbit. Two globular clusters that are near enough to each other to avoid being separated by the Galactic tidal field will be too close to avoid being disrupted by the mutual tidal field of the clusters. Alternately, if the inter-cluster separation is large enough that mutual gravitation is insufficient to cause the two clusters to merge into a single object then the effects of the Galactic tidal field would be great enough to cause the two clusters to move away from each other. Galactic globular clusters are among the oldest objects in the Galaxy so a primordial binary globular cluster will have undergone several Galactic orbits and thus could not survive, as a binary cluster, to the present day. The mass of M31 is approximately twice that of the Galaxy so it is even less likely that a binary globular cluster could survive in M31. A transient binary globular cluster, however, may be observed if two clusters passed near enough to interact with each other, although such an encounter would likely destroy the two clusters since a typical encounter timescale is longer than the time required for stars in one cluster to respond to the other cluster's gravitational field.

[1]Visiting Astronomer, Canada–France–Hawai'i Telescope (CFHT), operated by the National Research Council of Canada, le Centre National de la Recherche Scientifique de France, and the University of Hawai'i.

There is evidence that star clusters form in groups where the individual clusters are not gravitationally bound to each other. Lyngå & Wramdemark (1984) cite similarities in metallicities, stellar content, and stellar ages to argue that several multiple open clusters in the direction of the Gould Belt have a common origin. Such systems, however, are probably not dynamically bound binary clusters but simply clusters that have a common origin and are in the process of either separating or merging. The Galactic open clusters h & $\chi$ Persei are one possible such pair. This scenario is unlikely among the M31 globular clusters since that system appears to have an age comparable with that of the Galactic globular cluster system (Frogel et al. 1980, Bohlin et al. 1993).

The Large Magellanic Cloud (LMC), however, presents a more promising enviroment for binary clusters. The tidal field of the LMC is weak enough that a binary globular cluster could survive indefinitely (Innanen et al. 1972) without experiencing fatal disruption. Bhatia & Hatzidimitriou (1988) have catalogued 69 pairs of stellar clusters in the LMC with inter-cluster separations of less than 13″ ($\simeq$ 3.16 pc) and use statistical arguments to show that only half of these pairs are due to chance. Radial velocity studies have shown that the individual components of some of these multiple objects are interacting with each other while color–magnitude studies show that several of these multiple clusters are young compared to the ages of Galactic globular clusters. Because of their young age the LMC multiple clusters may not be true dynamically bound multiple cluster systems but rather clusters sharing common origins that are currently parting company.

G185 (Sargent et al. 1977) is a bright globular cluster located approximately $1\overset{'}{.}7$ south of the center of M31. We have identified a small object 4″ northwest of G185. This object has a surface brightness profile that is more extended then a stellar profile and is consistent with that of a loose globular cluster.

## 2. Observations and Data Reduction

We obtained several Johnson $V$- and Cousins $I$-band images of a $2' \times 2'$ field near the core of M31 that includes G185 ($\alpha_{2000} = 00^{\rm h}42^{\rm m}44\overset{\rm s}{.}2$, $\delta_{2000} = +41°14'28''$) and van den Bergh 2 (van den

Bergh 1969) ($\alpha_{2000}$ = 00$^{h}$42$^{m}$41$^{s}$.1, $\delta_{2000}$ = +41°15′26″; hereafter refered to as vdB2) as part of a program to obtain structural parameters for M31 globular clusters. These images were taken at the CFHT using the DAO/CFHT High Resolution Camera (HRCam) and the SAIC 1 CCD detector on the night of August 16/17, 1990 under photometric conditions. This CCD has a read-out noise of 6.5 $e^-$, a gain of 1.6 $e^-$ per analog-to-digital unit (ADU) and, in this configuration, an image scale of 0.″131 per pixel. The typical stellar full-width half-maximum (FWHM) for our data was ∼0.″7.

Bias subtraction and flat-fielding were performed in the standard manner. A full description of the observations and preprocessing is given in Holland et al. (1995) and a summary of the observations of the G185 field is given in Table 1. Due to the short exposure times only small numbers of cosmic ray events were detected on each frame. The IRAF[2] task NOAO.IMRED.CCDRED.COSMICRAY was used to identify and correct these contaminated pixels. The six $I$-band images were reregistered and averaged to improve the signal-to-noise ratio of the data. The calibrations (standard star reductions were taken from Hodder 1995) are

$$\mu_I = (19.4126 \pm 0.0270) - 2.5 \log_{10}(C_I), \qquad (1)$$

and

$$\mu_V = (19.6087 \pm 0.0046) - 2.5 \log_{10}(C_V), \qquad (2)$$

where the surface brightness, $\mu$, is in magnitudes/□″ and the observed count rate, $C$, is in ADUs. The gains for the fully preprocessed images were 9.6 $e^-$/ADU for the $I$-band image and 1.6 $e^-$/ADU for the $V$-band image.

The data reduction is similar to that described in Holland et al. (1995). A two-dimensional cubic spline was fit to and subtracted from each frame to remove the unresolved light from M31. The residual variations in the background were approximately 1%, not including small-scale variations due to dust lanes and surface brightness fluctuations arising from unresolved stars in the bulge and halo of M31. Figure 1 shows the $I$-band image of the field containing the three globular clusters. Figure 2 shows the same field with the unresolved light from M31 removed. Dust lanes and surface brightness fluctuations are clearly visible. Primini et al. (1993) report two *ROSAT* X-ray sources located within 0.′5 of G185 (sources 39 and 46) but find no evidence for an X-ray source in G185. We were unable to identify any unusual features within the error ellipses of the X-ray sources.

The IRAF implementation of the STSDAS task ELLIPSE was used to fit elliptical isophotes to each cluster. Surface brightness profiles were then extracted along the effective radius axes[3] of each cluster.

---

[2]Image Reduction and Analysis Facility (IRAF), a software system distributed by the National Optical Astronomy Observatories (NOAO).

The double cluster was fitted iteratively as follows. First, the fainter component (B) was fitted and and subtracted from the original image. Next, the brighter (A) component was fitted using the image with the B component subtracted. The A component was then subtracted from the original image and a new fit made to the B component. This cycle was repeated until the fitted ellipses for the two components were stable from one iteration to the next. Two or three iterations were sufficient for both the $V$- and $I$-band images. Figure 3 shows the subtractions of each component of G185 from the original image.

Describing the stellar point-spread-function (PSF) for each image was somewhat difficult. We attempted to fit both one-dimensional Moffatians (Moffat 1969) and one-dimensional multi-Gaussians (Bendinelli et al. 1987) to the stars on each frame but found that the wings of the PSFs (beyond $\sim1''.25$) were not well fit by either functional form. Therefore, we used DAOPHOT II (Stetson 1987, Stetson et al. 1990, Stetson 1991, Stetson 1992) to define a PSF in each band. We created noiseless bright artificial stars and fit elliptical isophotes to these in the same manner as was done for the globular clusters. This worked well out to $\sim1''.25$ but beyond this the uncertainties in the intensities of the fitted ellipses became larger than the PSF intensities. The difficulty in describing the PSFs beyond $\sim1''.25$ is due to small-scale (a few pixels) fluctuations in the image background. Since the program field is located only ~360 pc from the center of M31 there are large numbers of unresolved bulge and halo stars in the line of sight of each pixel. The PSF stars are sitting atop these large surface brightness fluctuations which in turn contribute to the shape of the PSF. In the core of the PSF the stellar luminosity is large enough to dominate the shape of the PSF, however in the wings of the PSF surface brightness fluctuations in the field can significantly influence the shape of the PSF. The large signal-to-noise ratio of the $I$-band data results in larger surface brightness fluctuations in the $I$-band than in the $V$-band making the wings of the PSF harder to measure in $I$ than in $V$. This can be seen in the wings of the PSF profiles in Figure 4.

The observed surface brightness profiles for each cluster and the PSFs are shown in Figure 4. G185-B is clearly more extended than the PSF suggesting that it is not merely a foreground star. However, the outermost points of the surface brightness profile of G185-B actually cross the profile of the PSF. This, and the extreme truncation of the cluster profile, is likely an artifact of the uncertainties in measuring the surface brightness profiles of G185-B and the PSF. G185-B has only ~3% of the luminosity of G185-A or vdB2. This results in the uncertainties in the surface brightness profile of G185-B being correspondingly larger than the uncertainties in the profiles of G185-A and vdB2 at similar fractions of the central surface brightness. Further, fitting and subtracting the unresolved background light from M31, and removing G185-A from the images,

[3]The effective radius, $r_{\rm eff}$, of an ellipse with semi-major and semi-minor axes $a$ and $b$ respectively is defined by $\pi r_{\rm eff}^2 \equiv \pi ab$. The effective radius axis is either of the two axes along which the edge of the ellipse intersects the edge of a circle of radius $r_{\rm eff}$.

results in additional uncertainties in the surface brightness measurements.

## 3. Results

### 3.1. Probability that G185-A and G185-B are Line-of-Sight Objects

We estimate the probability that two globular clusters $1\farcm7$ from the core of M31 will have an observed separation of $4''$ by chance as follows. Crampton et al. (1988) find that the projected density of the M31 globular cluster system drops as $R^{1/4}$ with distance from the center of M31. This density relation gives a probability of such a chance alignment on a $2' \times 2'$ field as 0.002. There are, however, ~5.34 such fields within an annulus of width $2'$ centered $1\farcm7$ from the center of M31. Therefore, the probability that a randomly selected field $1\farcm7$ from the center of M31 will contain such a pair of globular clusters is ~0.01. This calculation assumes that there is no angular dependence in the distribution of globular clusters in M31. In reality the small number density of globular clusters in our annulus will result in Poisson fluctuations in the number of clusters observed in different fields located at the same distance from the center of M31. To account for this we used the method of Bhatia & Hatzidimitriou (1988) to estimate the probability that a $2' \times 2'$ field containing three globular clusters will contain two globular clusters separated by $4''$ or less. This gives a probability of 0.016 for a single field. The probability of finding such a pair within our annulus, then, is $0.09 \pm 0.03$. A chi-square test indicates that this is significant at the 99% (2-$\sigma$) confidence level, however the small sample size (three clusters) suggests that this result not be given too much weight.

### 3.2. Colors

The total integrated magnitudes for G185-A, G185-B, and vdB2 were determined using aperture photometry with uncertainties estimated from fluctuations in the flat portion of the curve of growth for each cluster. To test this method of determining the uncertainties in the integrated magnitudes we built a series of artificial globular clusters by adding artificial stars to our images using realistic luminosity and radial density profiles. The luminosity function used was that of 47 Tuc (Hesser et al. 1987) with the horizontal branch stars distributed uniformly over the horizontal part of the horizontal branch. In practice we found the exact form of the luminosity function did not affect our results. The radial stellar density profiles used were generated from the Michie–King (Michie 1963, King 1966) models that we fit to the the clusters (see Sec. 3.3). From these artificial globular clusters we estimated the uncertainty in our aperture magnitudes to be less than 0.1 magnitudes. We also computed total magnitudes for each cluster by integrating under the observed surface brightness profiles. The two methods of measuring magnitudes agreed to within 1-$\sigma$ for all three clusters. The total calibrated aperture magnitudes are listed in Table 2. No corrections for reddening have been applied to these values.

Burstein & Heiles (1984) quote an external reddening of $E_{B-V} = 0.080 \pm 0.003$ for M31. Using $E_{V-I} = 1.36 E_{B-V}$ (Taylor 1986, Fahlman et al. 1989), and ignoring internal reddening within M31, gives $E_{V-I} = 0.11 \pm 0.01$ so all three globular clusters have dereddened colors that are consistent with the $(V-I)_0$ colors of the Galactic globular clusters (e.g. Peterson 1993). The $(V-I)$ colors of G185-A and G185-B suggest that they are old objects similar in age to the rest of the M31 globular cluster system. Therefore it is unlikely that G185-A and G185-B are a pair of young clusters that formed together and are currently separating. The field around G185 shows a large number of dust clouds which obscure the bulge of M31. The fact that none of the three clusters exhibits an unusual degree of reddening suggests that they all lie on the near side of the center of M31 and may be situated in front of the majority of the material in M31.

### 3.3. Cluster Concentrations

It can be shown (e.g. Holland et al. (1995) that convolving two one-dimensional surface brightness profiles is not equivalent to convolving two two-dimensional distributions then extracting a surface brightness profile. Fitting one-dimensional seeing-convolved Michie–King models overestimates the core radius, $r_c$, and tidal radius, $r_t$, by an amount that depends on the true values of these quantities. Therefore, we have modelled G185-A, G185-B and vdB2 as collections of scaled PSFs with realistic luminosity functions and isotropic Michie–King surface density distributions. Surface brightness profiles were derived from these model clusters and compared to the observed surface brightness profiles of the real clusters. To constrain the avaliable parameter space we restricted the allowable tidal radii of the models as follows. An upper limit for the tidal radius was determined by fitting one-dimensional isotropic seeing-convolved Michie–King models to each cluster. The lower limit for the tidal radius was taken to be the outer-most point of each observed surface brightness profile. Table 3 lists Michie–King parameters for each cluster and bandpass derived using two-dimensional modelling. The quoted uncertainties were estimated as follows. We built a series of artificial globular clusters with central potentials $W_0 = 5$, $r_c = 0\farcs5$, $r_t = 5\farcs35$, and ellipticities $\epsilon = 0.0$ then fit two-dimensional models to these clusters. The scatter in the recovered parameters were taken to be indicative of the uncertainties in our fitted models. This, however, does not include uncertainties due to our model PSF. We were unable to constrain the Michie–King parameters of G185-B as well as we were able to for the other two clusters. This is reflected in the larger uncertainties quoted for G185-B in Table 3.

G185-B appears to be a very loose cluster with a concentration of $0.67 \pm 0.17$. This is less than that of the majority of the Galactic globular clusters. In fact, only a handful of outer halo Galactic clusters have similar concentrations (Trager et al. 1993). This, and the fact that G185-B does not appear to be unusually reddened, argues that G185-B is not situated near the center of M31 and may be located well in front of M31. G185-B has an apparent tidal radius approximately three-and-a-half times smaller than that of G185-A. For this difference in size to be a purely geometric effect G185-B would have to be located at ∼3.5 times the distance of G185-A meaning that either G185-A or G185-B is not a member of the M31 globular cluster system. Huchra et al. (1991) give a radial velocity for G185 of $v_h = -483 \pm 25$ km s$^{-1}$, consistent with G185 being a member of the M31 globular cluster system. Since G185-A contributes ∼97% of the total light this radial velocity will be a reasonable estimate of the radial velocity of G185-A. The color of G185-B is not unduly red for an old globular cluster so it is unlikely that G185-B is being viewed through the bulge of M31. This suggests that G185-B is a member of the M31 globular cluster system and is intrinsically smaller than G185-A.

Our fitted core radius for G185-A is somewhat smaller than that found by Cohen & Freeman (1991). We find a core radius of $0\farcs46 \pm 0\farcs04$ ($= 1.6 \pm 0.1$ pc) and a tidal radius of $6\farcs13 \pm 0\farcs69$ ($= 21.5 \pm 2.4$ pc) where Cohen & Freeman (1991) find $r_c = 2.9$ pc and $r_t = 20$ pc. The difference in core radius is probably due to the improved seeing conditions of our observations.

### 3.4. Ellipticities

Figure 5 shows observed isophotes and some fitted isophotal ellipses for G185-A and G185-B. Observed weighted mean ellipticities, $\overline{\epsilon}$, and position angles, $\overline{\theta}_0$, were calculated for each cluster based on the fitted isophotes. These are presented in Table 4 along with the number, $N$, of fitted ellipses used to compute these values. Figures 6 and 7 show the radial variations of these quantities for G185-A, G185-B and the PSF. Position angles are measured from north to east on the sky and the quoted uncertainties are due to the scatter in fitted position angle from the center to the outer edge of each cluster, and the estimated uncertainty in orienting the CCD images with respect to north. Large uncertainties are indicative of a large radial variation in position angle. Figures 6 and 7 show that radial changes in the ellipticity and orientation of the PSF are echoed in the shape of G185-B suggesting that seeing is the dominant effect in determining the shape of G185-B.

The observed ellipticity of a semi-resolved globular cluster is dominated by the shape of the PSF out to ∼8 times the seeing FWHM Holland et al. (1995). This corresponds to $5\farcs7$ in the $I$-band and $6\farcs9$ in the $V$-band. However Figures 6 and 7 show that ellipses could only be fit out to ∼5″ for the best defined cluster. This, and the similarities between the ellipticities of the clusters and the PSFs, suggests that the observed projected ellipticities of G185-A and G185-B are due to seeing effects.

To test this conclusion we built a series of artificial globular clusters based on G185-A but with ellipticities between 0.00 and 0.21. These were placed on the $I$-band data frame and elliptical isophotes fitted to them as was done for the real globular cluster data. The recovered $I$-band ellipticities at a given location on the image showed no trend with input ellipticity and had a mean of $\overline{\epsilon} = 0.036 \pm 0.004$ (standard deviation). We added artificial clusters to three locations on the original data frame and found that the recovered mean ellipticities and position angles varied significantly with position. Recovered ellipticities were between 0.029 and 0.041 while recovered position angles varied by up to 60°. We believe this positional dependence is due to the presence of large pixel-to-pixel surface brightness fluctuations in the unresolved light from the bulge of M31. These fluctuations are blurred by the seeing so that they have the same shape and orientation as the PSF. The luminosity of a fluctuation depends on the number of stars in the line of sight and thus its location on the frame relative the the center of M31. Further, the PSF on an HRCam image is known to vary with position on the frame so the size and shape of the surface brightness fluctuations will also vary with position. This would account for the variation in recovered ellipticity and position angle with location on the frame and partially explain why the three clusters are observed to have different ellipticities and position angles. A further reason for cluster-to-cluster ellipticity variations is that the light from each cluster is dominated by a small number of stars near the tip of the red giant branch which will result in small-number statistics dominating the observed shapes of the clusters. This effect should be more noticable in small clusters and may explain the large observed ellipticity of G185-B. This conclusion is supported by the fact that the orientation of G185-B is significantly different in each color. The fact that the $V$- and $I$-band ellipticities of G185-B are similar but the orientations are different suggests that G185-B is not simply a double star.

The apparent elongation of G185-A has the cluster pointing 13.°1 west of G185-B. However, this is nearly the same direction that the PSF is oriented, and G185-B is not distorted in the direction of G185-A. We were able to reproduce the observed $V$-band shape of G185-A with artificial clusters of zero ellipticity but we were unable to reproduce its observed ellipticity in the $I$-band. It is, therefore, uncertain if G185-A is actually oriented towards G185-B or the observed elongation is merely due to the seeing and background fluctuations. As we noted in Sec. 2 the surface brightness fluctuations in the $I$-band are greater than those in the $V$-band so we would expect them to cause a greater distortion in the observed shape of G185-A in the $I$-band image.

### 3.5. The Roche Limit of the System

The mean radial velocity of a globular cluster in the M31 system relative to M31 is $v = 125$ km s$^{-1}$ (Huchra et al. 1991) which implies an encounter timescale of at least $10^5 - 10^6$ years, assuming a near head-on encounter. This minimum encounter time is similar to the crossing time for a star in a globular cluster so interacting clusters will have sufficient time to be spatially distorted by each others' tidal fields. We have calculated the shapes of the Roche lobes for G185-A and G185-B assuming that the two clusters are interacting and that they can be modelled by point masses. In reality the two clusters are not true point sources, but since the half-mass radii of the Michie–King models fitted to them are comparable to their fitted core radii, this assumption is reasonable. The fitted half-mass radii for G185-A and G185-B respectively are 1.″05 ± 0.″07 and 0.″49 ± 0.″16.

Figure 8a shows that the cut-offs imposed by the tidal field of M31 are approximately the same as the cut-offs imposed by the mutual tidal fields of the two clusters if G185-A and G185-B have a true separation of ~35 pc ($\simeq 10''$). If the true separation is greater than this then the shapes of the clusters will be determined by the tidal field of M31. If, however, the true separation is smaller then inter-cluster tidal forces will have a significant effect on the internal dynamics of each cluster. Figure 8b shows that if the true separation is equal to the observed projected separation ($4'' \simeq 14$ pc) then the cut-offs imposed by their mutual tidal fields are closer to the clusters' centers than the tidal cut-offs imposed by M31. Since we observe light beyond these Roche limits the clusters are either not interacting or have not had time to dynamically respond to each others' tidal fields.

## 4. Conclusions

G185-A and G185-B appear to be two separate globular star clusters in M31. There is no direct evidence that the two clusters are interacting. We find no evidence of tidal disruption in either clusters although the $I$-band ellipticity of G185-A is greater than can be accounted for due to seeing effects and G185-A is approximately oriented towards G185-B.

This research was supported by operating grants from the Natural Sciences and Engineering Research Council of Canada and has made use of the NASA/IPAC Extragalactic Database (NED) which is operated by the Jet Propulsion Laboratory, Caltech, under contract with the National Aeronautics and Space Administration.

## REFERENCES

Bendinelli, O., Parmaggiani, G., Piccioni, A., & Zavatti, F. 1987, AJ, 94, 1095

Bhatia, R. K., & Hatzidimitriou, D. 1988, MNRAS, 230, 215

Bohlin, R. C., Deutsch, E. W., McQuade, K. A., Hill, J, K., Landsman, W. B., O'Connell, R. W., Roberts, M. S., Smith, A. M., & Stecher, T. P. 1993, ApJ, 417, 127

Burstein, D., & Heiles, C. 1984, ApJS, 54, 33

Cohen, J. G., & Freeman, K. C. 1991, AJ, 101, 483

Crampton, D., Cowley, A. P., Schade, D., & Chayer, P. 1985, ApJ, 288, 494

Fahlman, G. G., Richer, H. B., Searle, L., & Thompson, I. B. 1989, ApJ, 343, L49

Frogel, J. A., Persson, S. E., & Cohen, J. G. 1980, ApJ, 240, 785

Hesser, J. E., Harris, W. E., VandenBerg, D. A., Allwright, J. W. B., Shott, P., & Stetson, P. B. 1987, PASP, 99, 739

Hodder, P. J. C. 1995, Ph.D. Thesis, UBC, in preparation

Holland, S., Fahlman, G. G., & Richer, H. B. 1995, in preparation

Huchra, J. P., Brodie, J. P., & Kent, S. M. 1991, ApJ, 370, 495

Innanen, K. A., Wright, A. E., House, F. C., & Keenan, D. 1972, MNRAS, 160, 249

King, I. R. 1966, AJ, 71, 64

Lyngå, G., & Wramdemark, S. 1984, A&A, 132, 58

Michie, R. W. 1963, MNRAS, 125, 127

Moffat, A. F. J. 1969, A&A, 3, 455

Peterson, C. 1993, in Structure and Dynamics of Globular Clusters, eds. S. Djorgovski, & G. Meylan, (San Francisco : ASP), ASP Conf. Ser. 50, p 337

Primini, F. A., Forman, W., & Jones, C. 1993, ApJ, 410, 615

Sargent, W. L. W., Kowal, C. T., Hartwick, F. D. A., & van den Bergh, S. 1977, AJ, 82, 947

Stetson, P. B. 1987, PASP, 99, 191

Stetson, P. B. 1991, in Third ESO/ST-ECF Data Analysis Workshop, eds. P. J. Grosbøl & H. R. Warmels, (Garching : ESO), p 187

Stetson, P. B. 1992, in Astronomical Data Analysis Software and Systems I, eds. D. M. Worrall, C. Biemesderfer, & J. Barnes, (San Francisco : ASP), ASP Conf. Ser., 25, p 297

Stetson, P. B., Davis, L. E., & Crabtree, D. R. 1990, in CCDs in Astronomy, ed. G. H. Jacoby, (San Francisco : ASP), ASP Conf. Ser., 8, p 289

Taylor, B. J. 1986, ApJS, 60, 577

Trager S. C., Djorgovski, S., & King, I. R. 1993, in Structure and Dynamics of Globular Clusters, eds. S. Djorgovski, & G. Meylan, (San Francisco : ASP), ASP Conf. Ser. 50, p 347

van den Bergh, S. 1969, ApJS, 19, 145

---

TABLE 1. Log of the Observations.

| Frame | Filter | UT | Exposure (s) | Airmass | FWHM |
|---|---|---|---|---|---|
| 83924 | $I$ | 12:24:54.1 | 100 | 1.099 | $0\farcs68$ |
| 83925 | $I$ | 12:31:03.4 | 100 | 1.094 | $0\farcs83$ |
| 83926 | $I$ | 12:34:32.7 | 100 | 1.091 | $0\farcs72$ |
| 83927 | $I$ | 12:38:06.0 | 100 | 1.089 | $0\farcs68$ |
| 83928 | $I$ | 12:41:37.4 | 100 | 1.086 | $0\farcs69$ |
| 83929 | $I$ | 12:45:09.7 | 100 | 1.084 | $0\farcs70$ |
| 83922 | $V$ | 12:15:21.7 | 200 | 1.109 | $0\farcs69$ |

TABLE 2. Integrated Magnitudes.

| Cluster | $I$ | $(V-I)$ |
|---|---|---|
| G185-A | $13.05 \pm 0.03$ | $1.19 \pm 0.03$ |
| G185-B | $16.87 \pm 0.10$ | $1.09 \pm 0.14$ |
| vdB2 | $13.35 \pm 0.03$ | $1.17 \pm 0.03$ |

TABLE 3. Michie–King Model Fits.

| Cluster | Filter | $W_0$ | $r_c$ | $r_t$ | $c$ |
|---|---|---|---|---|---|
| G185-A | $I$ | $5.5 \pm 0.3$ | $0\farcs46 \pm 0\farcs03$ | $6\farcs26 \pm 1\farcs09$ | $1.14 \pm 0.07$ |
| | $V$ | $5.2 \pm 0.4$ | $0\farcs50 \pm 0\farcs08$ | $6\farcs09 \pm 0\farcs58$ | $1.07 \pm 0.09$ |
| G185-B | $I$ | $3.0 \pm 1.0$ | $0\farcs39 \pm 0\farcs13$ | $1\farcs70 \pm 0\farcs25$ | $0.67 \pm 0.17$ |
| | $V$ | $3.0 \pm 1.0$ | $0\farcs40 \pm 0\farcs13$ | $1\farcs77 \pm 0\farcs30$ | $0.67 \pm 0.17$ |
| vdB2 | $I$ | $7.0 \pm 0.3$ | $0\farcs17 \pm 0\farcs02$ | $5\farcs90 \pm 0\farcs67$ | $1.53 \pm 0.09$ |
| | $V$ | $6.5 \pm 0.3$ | $0\farcs13 \pm 0\farcs02$ | $5\farcs90 \pm 0\farcs67$ | $1.39 \pm 0.09$ |

TABLE 4. Ellipticities and Position Angles.

| Cluster | Filter | $\bar{\epsilon}$ | $\overline{\theta}_0$ | N |
|---|---|---|---|---|
| G185-A | $I$ | 0.105 ± 0.005 | $-73\overset{\circ}{.}8 \pm 4\overset{\circ}{.}3$ | 34 |
| | $V$ | 0.034 ± 0.004 | $-87\overset{\circ}{.}1 \pm 38\overset{\circ}{.}7$ | 29 |
| G185-B | $I$ | 0.108 ± 0.012 | $-52\overset{\circ}{.}8 \pm 23\overset{\circ}{.}8$ | 19 |
| | $V$ | 0.115 ± 0.016 | $-14\overset{\circ}{.}4 \pm 14\overset{\circ}{.}5$ | 16 |
| vdB2 | $I$ | 0.036 ± 0.001 | $-78\overset{\circ}{.}9 \pm 5\overset{\circ}{.}2$ | 29 |
| | $V$ | 0.027 ± 0.002 | $-8\overset{\circ}{.}0 \pm 15\overset{\circ}{.}8$ | 26 |
| PSF | $I$ | 0.199 ± 0.006 | $-64\overset{\circ}{.}1 \pm 31\overset{\circ}{.}5$ | 23 |
| | $V$ | 0.029 ± 0.002 | $-7\overset{\circ}{.}8 \pm 33\overset{\circ}{.}4$ | 24 |

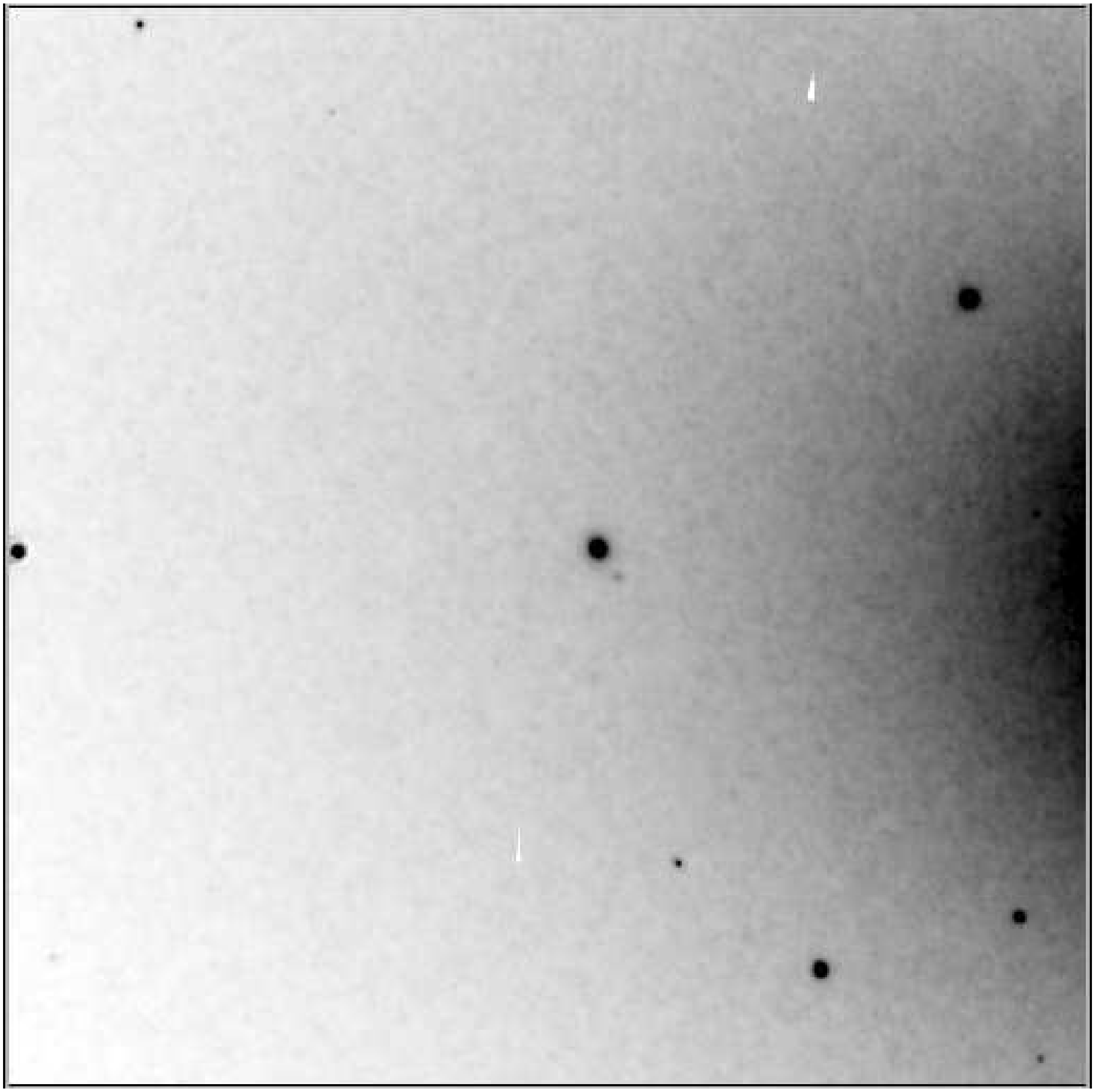

Fig. 1.— This plate shows the $I$-band image of G185-A (center), G185-B (just below and to the right of G185-A), and vdB2 (upper right). The center of M31 is located to the right of the frame. The top of the image is oriented 84° east of north. The object above G185-A is an image cursor.

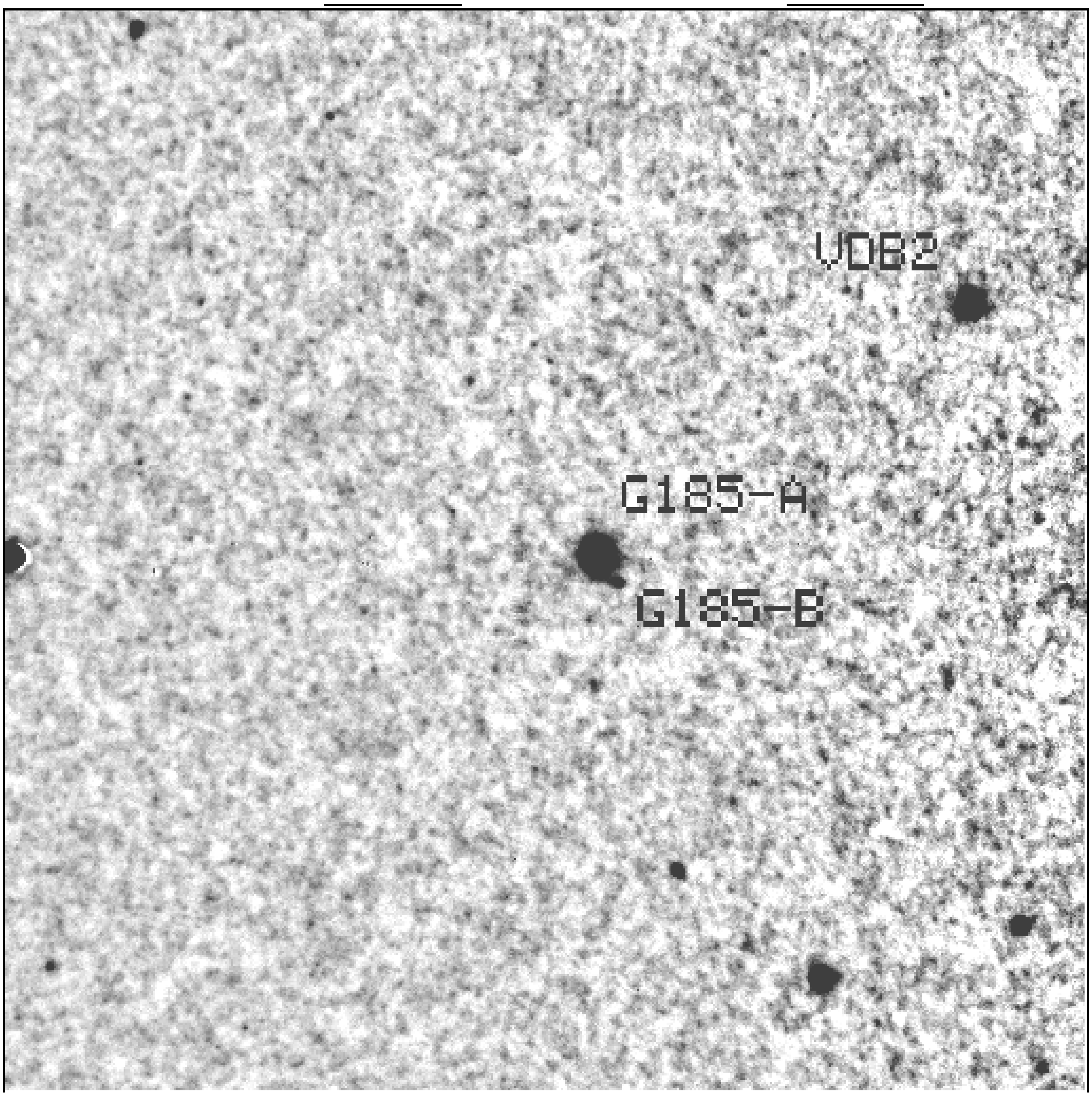

Fig. 2.— This plate is the same as Figure 1 except the unresolved light from the bulge of M31 has been subtracted leaving dust lanes and surface brightness fluctuations visible. Notice the surface brightness fluctuations are greater on the right-hand side of the image (towards the center of M31).

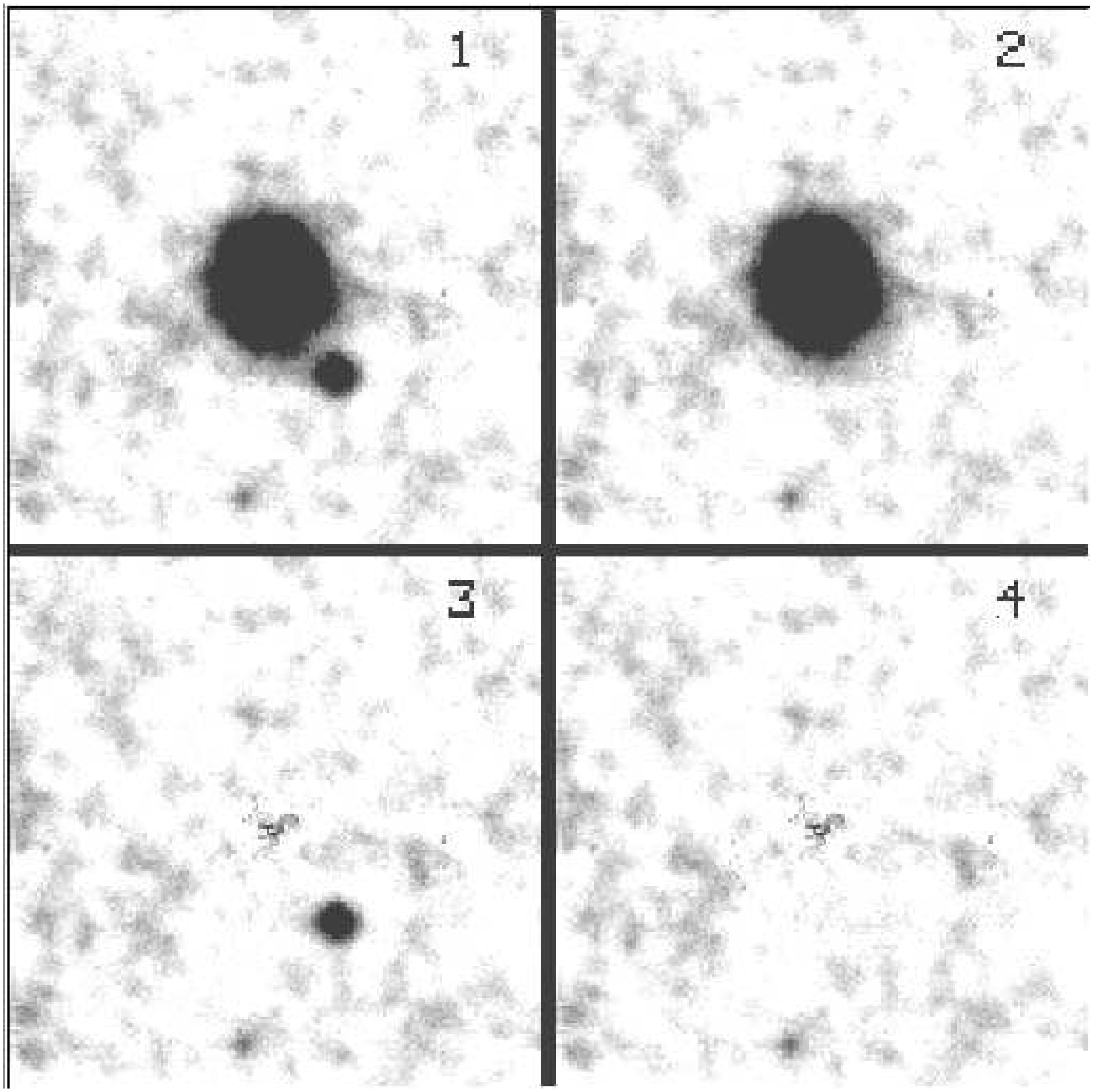

Fig. 3.— This plate shows G185-A and G185-B at various stages of the iterative fit-and-subtract procedure described in Sec. 2. (1) shows the two clusters G185-A and G185-B. (2) shows the same field with G185-B subtracted while (3) shows the field with only G185-A subtracted. (4) shows both clusters removed from the field.

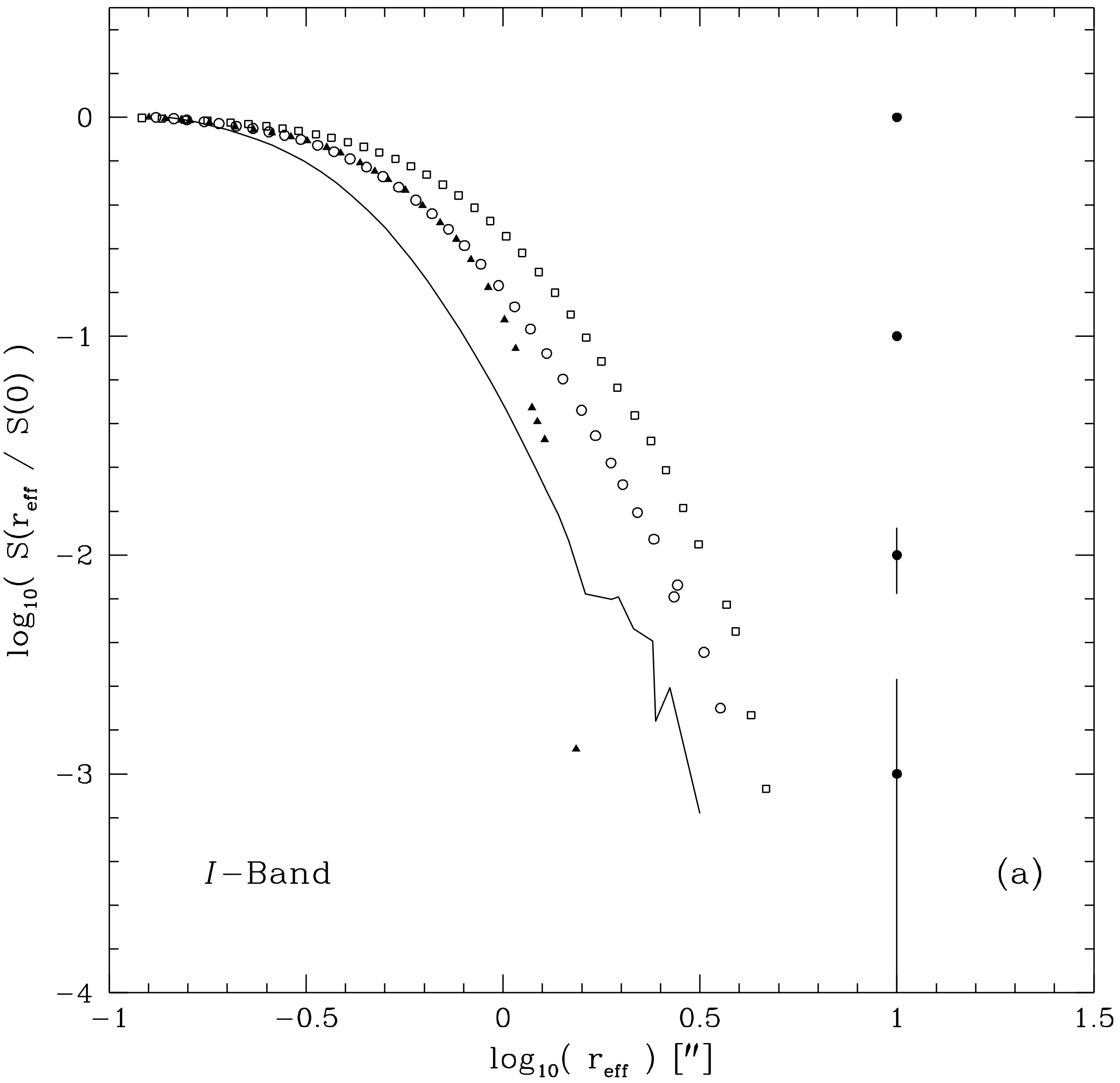

Fig. 4a.— The normalized observed surface brightness profiles of each cluster and the PSF in the $I$-band (a) and in the $V$-band (b). G185-A is denoted by open squares, G185-B by filled triangles, vdB2 by open circles and the PSF by the solid line. The noise in the wings of the $I$-band PSF is due the higher signal-to-noise ratio of the $I$-band image causing larger surface brightness fluctuations in the background. The error bars show the approximate uncertainty in the surface brightness profiles of G185-A and vdB2 at the given intensity levels. G185-B is nearly four magnitudes fainter than the other two clusters so its uncertainties will be greater than indicated by the error bars To estimate the uncertainty in the G185-B profiles shift the given error bars up by approximately 1.5 units.

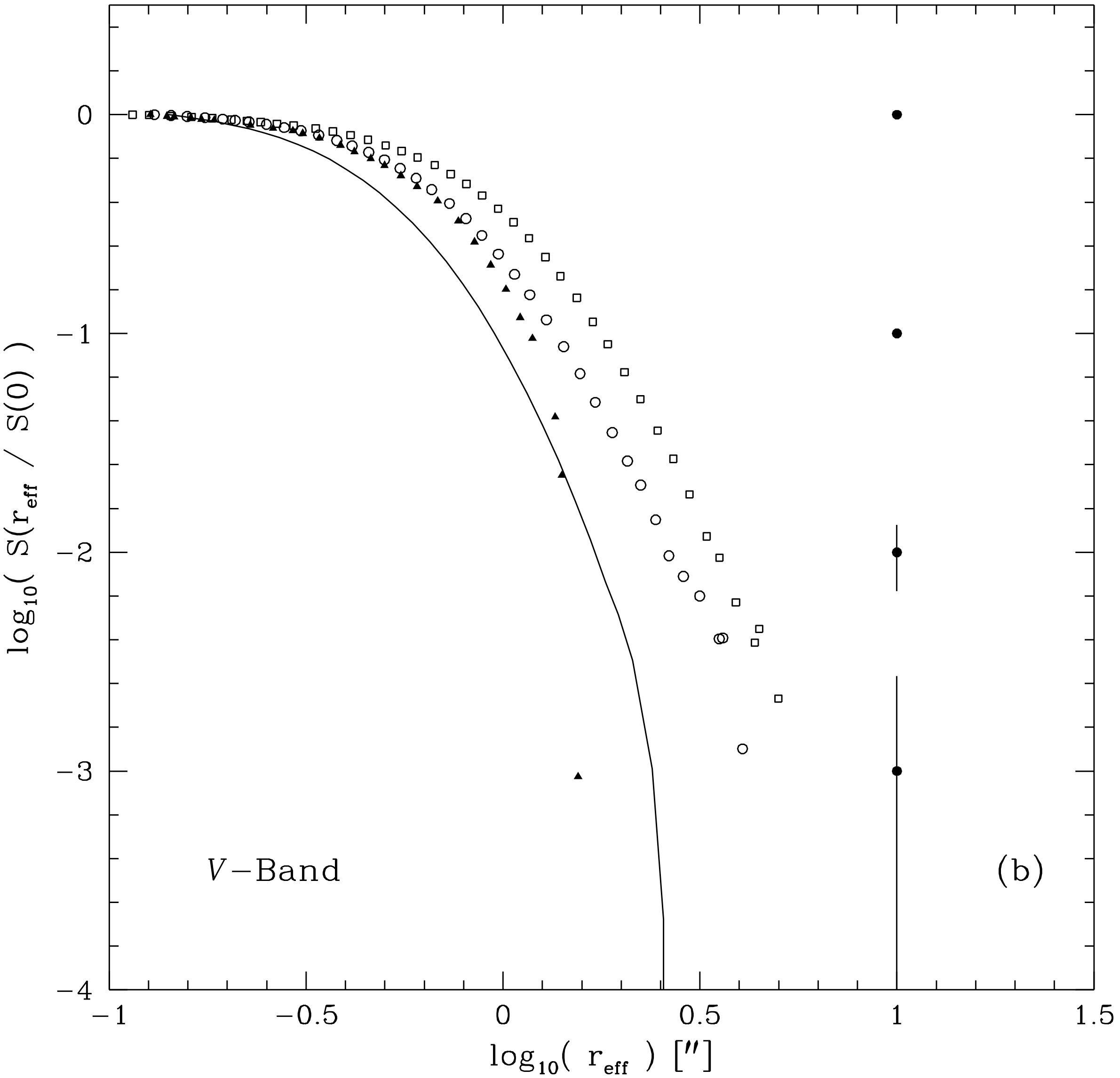

Fig. 4b.—

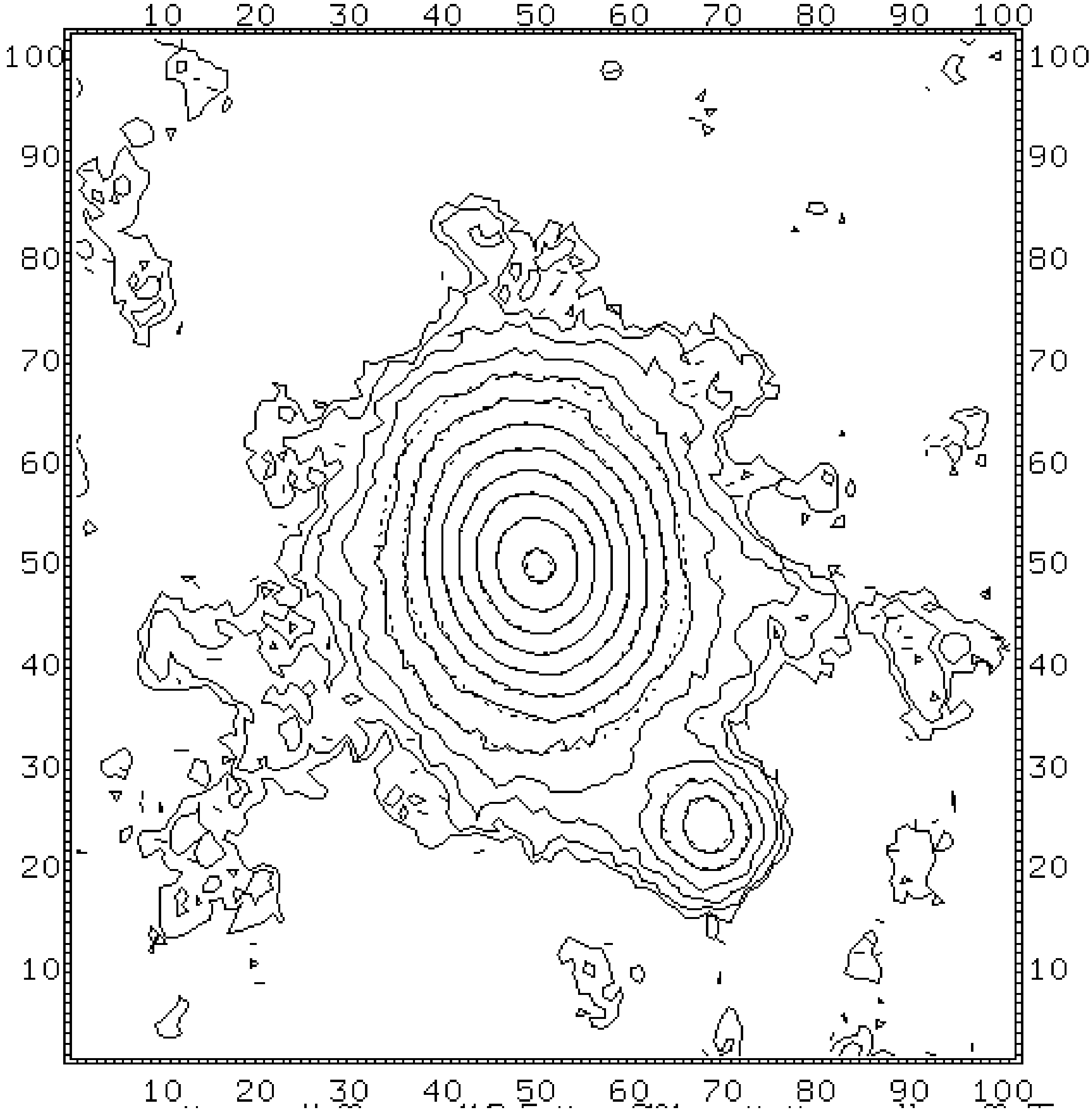

Fig. 5.— This figure shows a contour plot of the $I$-band images of G185-A and G185-B. The solid contours show surface brightnesses spaced between $\mu_I$ = 14.5 and $\mu_I$ = 20.0 while the dashed contours show fitted isophotal ellipses. G185-B is located 4″ northwest of G185-A. The center of M31 is approximately 6° west of north (the top of the plot is points towards 84° east of north).

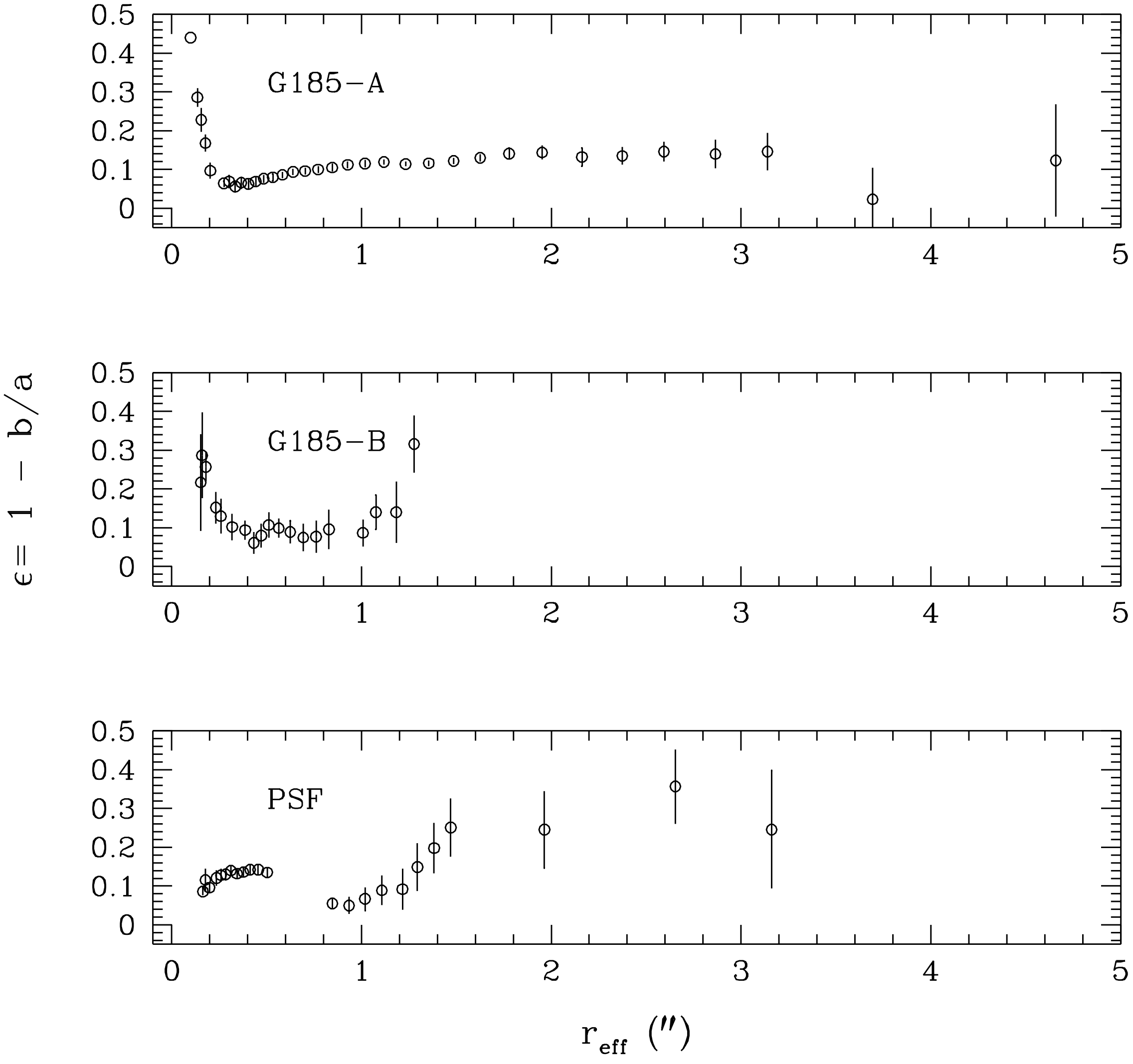

Fig. 6.— *I*-band ellipticity profiles of G185-A, G185-B, and the PSF. Only points where ELLIPSE (see Sec. 2) was able to successfully fit an elliptical isophote are shown. The increase in ellipticity in the inner 0.″5 is probably an artifact of the fitting algorithm and seeing effects. We are unable to reliably measure ellipticities or position angles in the outer regions of the clusters since Poisson noise and surface brightness fluctuations dominate the signal there.

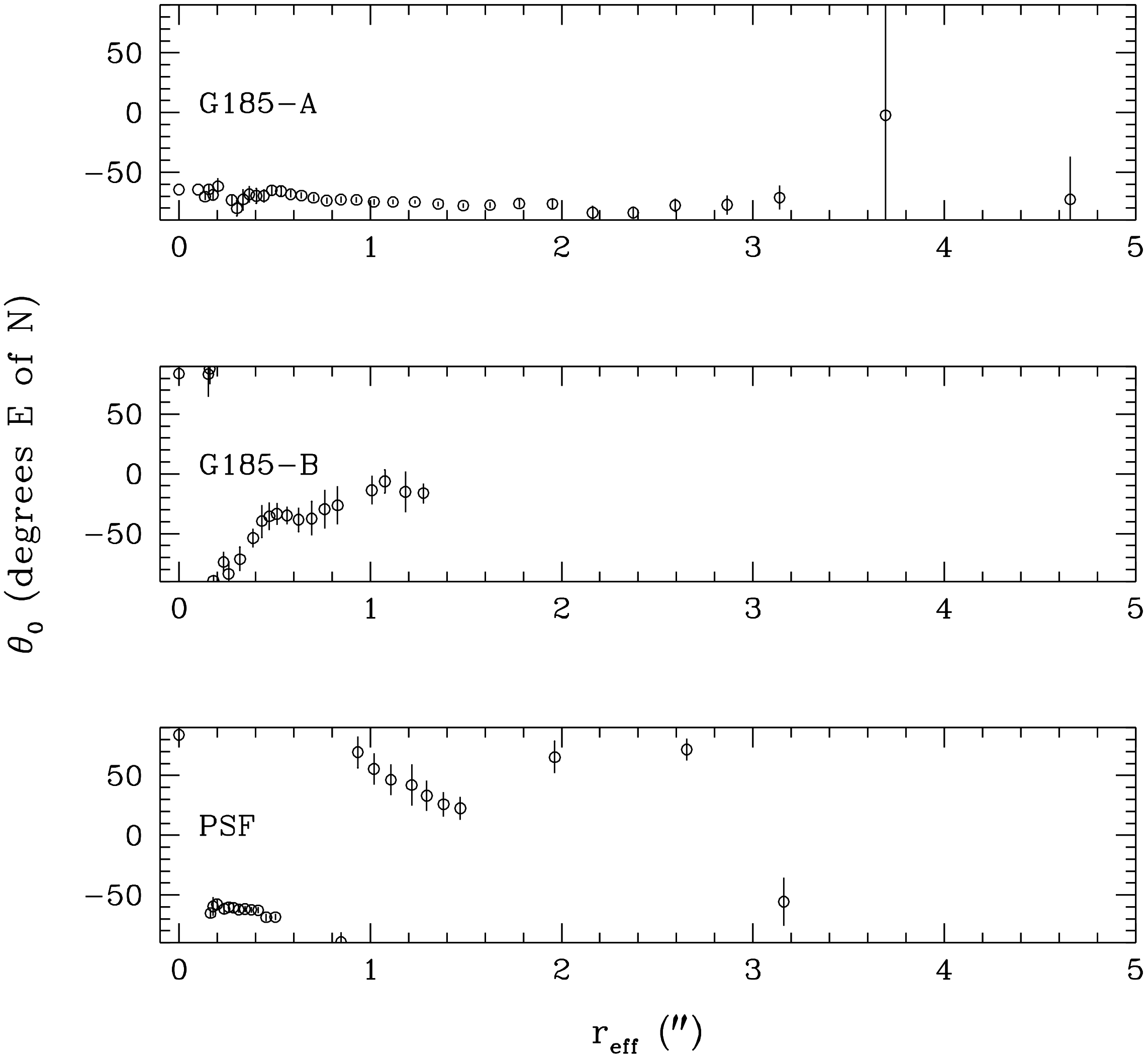

Fig. 7.— $I$-band position angle profiles of G185-A, G185-B, and the PSF. The same caveats apply as did for Figure 6. The position angles are measured in degrees east on the sky from north. Since these profiles are all measured within the inner regions of the clusters (where seeing dominates the observed shape, see Sec. 3.5) the large variations in $\theta_0$ for G185-B are probably due to the seeing and not intrinsic to the cluster.

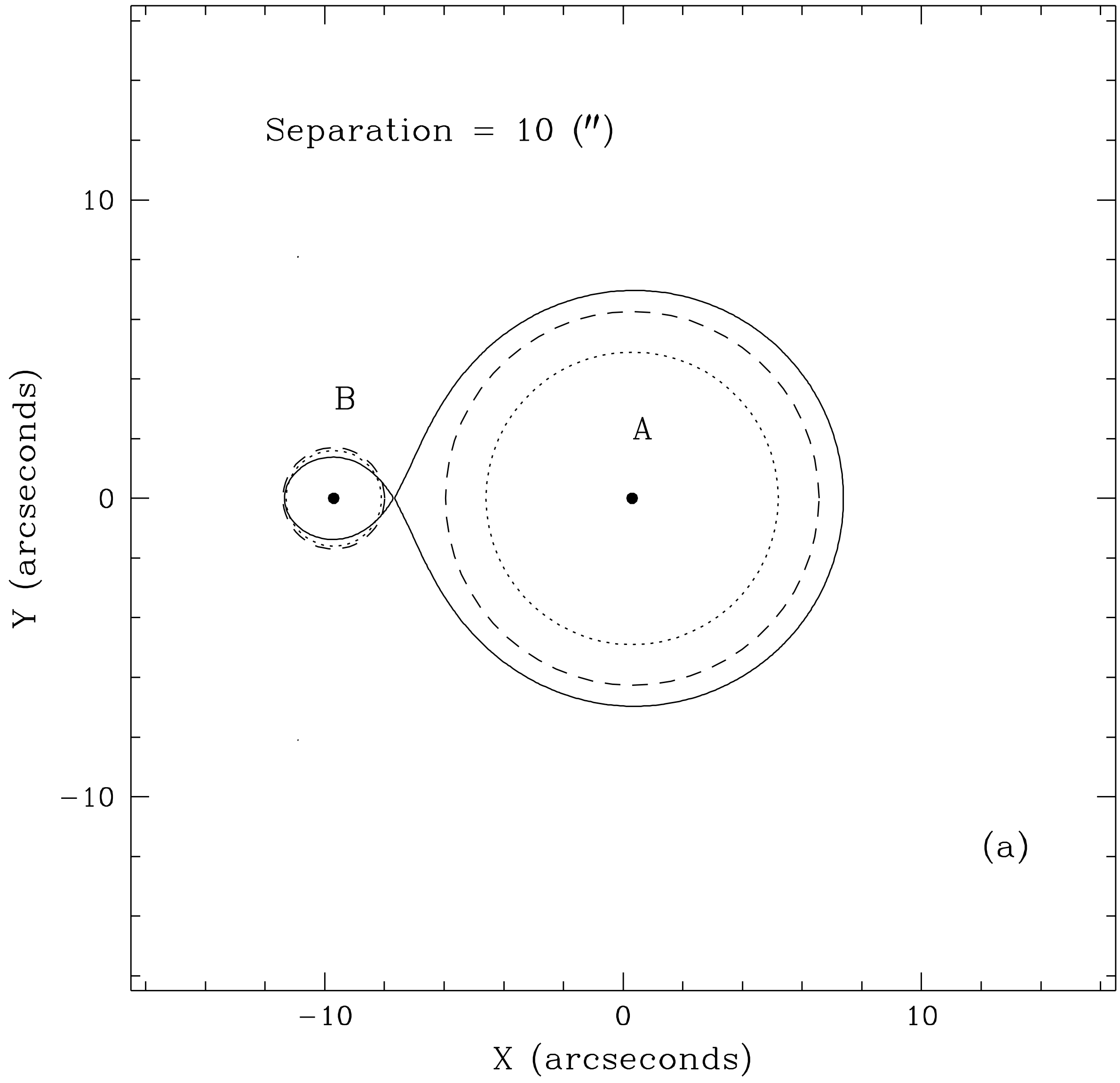

Fig. 8a.— Figures 8a & 8b show the Roche lobes for the two clusters computed assuming the clusters are interacting and can be approximated by point masses (see Sec/ 3.5 for a justification of the latter assumption). The solid lines show the Roche lobes while the dashed lines are the fitted $I$-band Michie–King tidal radii from Table 3. The dotted lines show the outermost point of the observed surface brightness profile for each cluster. The two clusters in Figure 8a have a true separation of 10″ ($\simeq$ 35 pc) while in Figure 8b they have a true separation of 4″ ($\simeq$ 14 pc, the same as the observed projected separation between G185-A and G185-B).

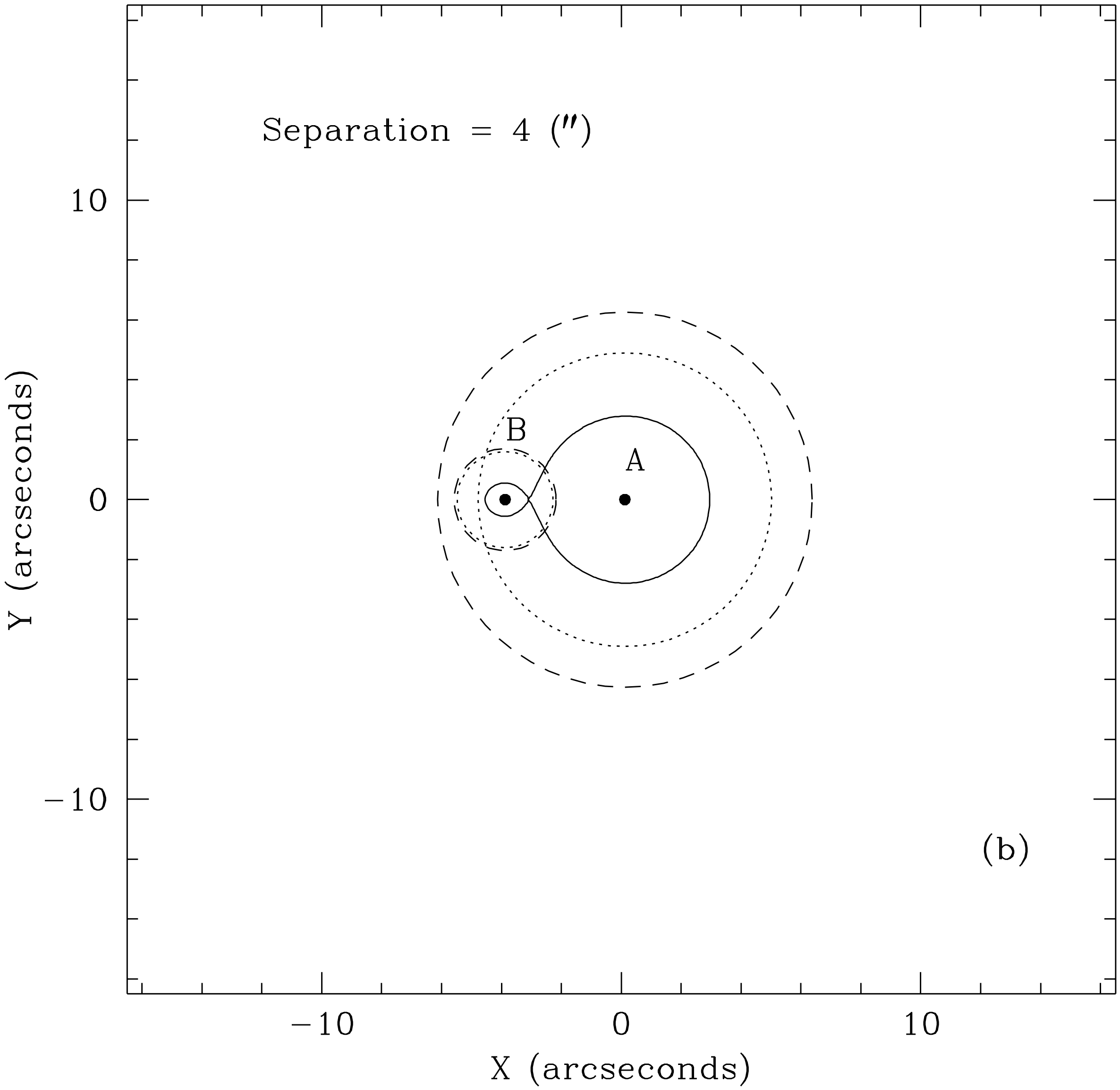

Fig. 8b.—